# Meta-level issues in Offloading: Scoping, Composition, Development, and their Automation

André DeHon, Hans Giesen, Nik Sultana, Yuanlong Xiao
University of Pennsylvania
USA

## ABSTRACT

This paper argues for an accelerator development toolchain that takes into account the whole system containing the accelerator. With whole-system visibility, the toolchain can better assist accelerator scoping and composition in the context of the expected workloads and intended performance objectives. Despite being focused on the 'meta-level' of accelerators, this would build on existing and ongoing DSLs and toolchains for accelerator design. Basing this on our experience in programmable networking and reconfigurable-hardware programming, we propose an integrative approach that relies on three activities: (i) generalizing the focus of acceleration to *offloading* to accommodate a broader variety of non-functional needs—such as security and power use—while using similar implementation approaches, (ii) discovering what to offload, and to what hardware, through semi-automated analysis of a whole system that might compose different offload choices that change over time, (iii) connecting with research and state-of-the-art approaches for using domain-specific languages (DSLs) and high-level synthesis (HLS) systems for custom offload development. We outline how this integration can drive new development tooling that accepts models of programs and resources to assist system designers through design-space exploration for the accelerated system.

## 1 THINKING OUTSIDE THE ACCELERATOR

Accelerators are typically well-defined, specialized subsystems whose usefulness depends on their contribution to a larger system. The usefulness of accelerators depends on considerations outside them—considerations at the level of the whole system and beyond, such as workload characteristics and system evolution, which can influence the extent to which a particular accelerator candidate satisfies the system needs.

A development environment for accelerators would be thus incomplete if it did not integrate design considerations that are external to the accelerator. The purpose of this paper is to explore some of these issues encountered when designing such an environment based on our experience in programmable networking and reconfigurable-hardware programming.

We start by observing the benign generalization of acceleration to *offloading*. Offload development uses similar techniques and tools as accelerator development but includes a broader set of requirements and objectives—such as reducing cost through reuse, reducing heat dissipation with specialized operators, reducing sensitive data exposure using specialized security hardware to safely store keys, and all of the above.

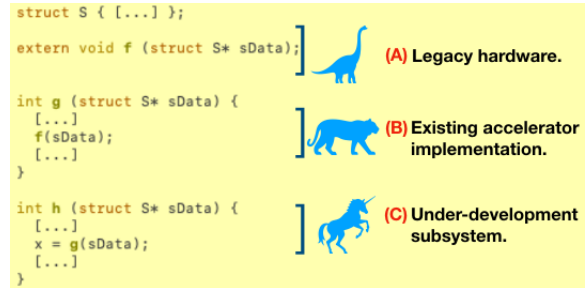

**Figure 1: Even if originating from the same codebase, different subsystems might be intended or eligible for mapping to different types of hardware.**

Next we focus on three meta-level issues that influence the usefulness of offload deployment. **(i)** *Scoping* involves finding parts of a system that would benefit from offloading. This involves analyzing the system to identify subsystems and characterizing the non-functional dependencies between them—e.g., performance or security implications. For example, if the goal of offloading is acceleration, then we need to derive time-characteristics of the accelerator. Alternatively, the goal might be to conserve SRAM on a precious resource, trading it for cheaper memory at the cost of time overhead.

As done in Flightplan [8], a single system might offload different subsystems to different hardware to obtain non-functional benefits such as acceleration. A system might use a range of different implementation languages and their toolchains, but for simplicity Fig. 1 shows a single, C-like language to describe functionality that is **(A)** offloaded to third-party hardware with a well-understood behavior, **(B)** offloaded to hardware implementations whose behavior we can change, **(C)** potentially offloaded to in-development hardware whose behavior is still being defined—scoping helps define the characteristics of that implementation and its behavior.

**(ii)** *Composition* assembles offloads into the system, such as **(A-C)** in Fig. 1. In our work we reason about compositions through models that capture subsystem profiles across heterogeneous hardware (§2.1), but it is an open question how to model different kinds of targets and how to structure models to reduce the complexity of reasoning about them. In our work, composition is inferred when control-flow crosses an annoted boundary in the code. In general the system and its offloaded subsystems can use a diverse mix of input languages and toolchains, including domain-specific approaches such as Dandelion [6] for LINQ queries or Emu [7] for packet processing. Accommodating this diversity requires reasoning about dependencies within a system and across subsystems to relate resource budgets with input workloads and output expectations.

**(iii)** *Development* supports exploring the design space *now*, feeding back into scoping and composition choices, and planning for the *future* as the system evolves. For example, this could explore the use of different targets for **(A-C)** to minimize both cost for the operator and end-to-end latency for end-users of a system. This involves understanding the sensitivity to input workloads based on model analysis or implementation profiling, for example, to understand scaling behavior. In our work we explored this in the context of planning (offload selection and provisioning) (§2.1).

Having introduced meta-level considerations, we say more on how these derive from our past work (§2) and sketch how they play a role in a cross-community research vision (§3).

## 2 TWO SETTINGS THAT USE OFFLOADING

### 2.1 Programmable networking

In our work on Flightplan [8] we offloaded fragments of P4 [1] packet processors to heterogeneous hardware. Fragments can invoke complex external logic such as link-layer Forward Error Correction [4]. For **scoping**, the programmer annotates the program and Flightplan explores all coarsenings of those code segments. It also helps reason about **composition** by capturing context to resume a computation on a different target. To help with **development**, it uses a planner that ingests program abstractions, information about the network and its hardware, and the operator's objectives, to produce a series of mappings to hardware as well as a performance, resources costs, and overhead forecast.

### 2.2 Reconfigurable hardware

In our work for reconfigurable hardware, we offload computational kernels to specialized, spatial operators to increase throughput and decrease latency. In this work the system is written in C and **composes** subprograms interconnected using streams [2, 3]. During **development** the designer progressively defines new subsystem implementations; these implementations are functionally equivalent, and earlier implementations are used to validate later implementations. This process feeds back to **scoping**.

## 3 VISION FOR OFFLOAD DEVELOPMENT

Developing frameworks that assist with both decomposition and refinement has value to designers and operators, but important open questions remain. We focus on two that we grappled with in our work: what models capture adequate information to automate reasoning about composition and hardware target selection (§3.2), and how could an offload development environment help with design-exploration involving meta-level issues such as scoping and composition (§3.1).

### 3.1 Models

Flightplan (§2.1) uses a *program model* to describe control- and data-flow, and delineate subsystem scopes. But it does not handle language features that exceed P4's expressiveness, such as those for asynchrony and concurrency that other programmable targets support. Being able to reason about these effects would increase the automation scope to more complex logic and hardware targets.

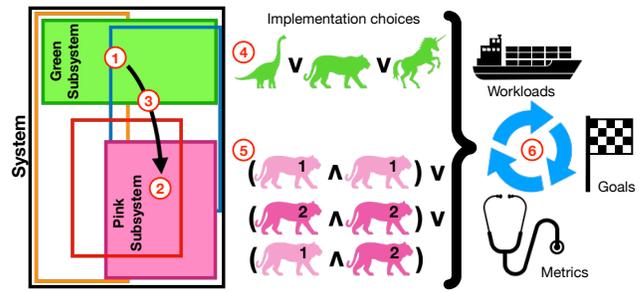

**Figure 2: Outline of an accelerator development framework.**

Flightplan also uses a *resource model* that captures features including latency, throughput, energy, and cost. But further work is needed on improve on this to: (i) accurately model sub-unit resources, such as FPGA resources and cores on a chip multiprocessor (CMP), and (ii) explore the trade-off of accuracy and tractability in cross-scale reasoning. For example when reasoning about congestion or concurrency on buses, reductionist modeling leads to a state-space explosion.

### 3.2 Exploration framework

Offload exploration is tedious and error-prone if done manually. Flightplan offers a proof-of-concept automation for exploration, but further generalization is needed. This includes exploring the spectrum between testbed experiments, simulations, and mathematical modeling to increase reasoning speed. Flightplan relies on testbed experiments to sample resource usage, but while these samples are high-fidelity, they take time and human effort to acquire. There is a need for tools to automate the characterization of resources and mappings, to lower the time and effort needed.

We envisage a workload-guided feedback process to identify offload candidates. Fig. 2 provides a sketch of such a system, based on our experience designing Flightplan. The design helps with scoping subsystems, such as the ① green and ② pink candidates, and reasoning about the ③ interfaces across intermediate subsystems.

An important design goal involves comparing alternative mappings to ④ different hardware types, and ⑤ different implementations on the same hardware. For flexibility, this process should work with different types of artefacts, including black-box implementations and abstract specifications of their execution profile.

Finally, ⑥ an evaluation process explores different compositions based on both quantitative and propositional reasoning. A search strategy (e.g, [5]) is used to spare designers and operators from time-consuming trial-and-error. This could be integrated with vendor toolchains to obtain feedback from proprietary quantifications of resource use on ASICs or reconfigurable hardware. Parts of this process could be seen as a "provisioning, allocation, scheduling, placement and routing" at a larger scale and involving coarser units of computation. Flightplan provides a starting point for such a system [8, §7.2.2] that we hope to generalize for more hardware targets and environments beyond programmable networking.

<mark>Meta-level issues in Offloading: Scoping, Composition, Development, and their Automation</mark> <mark>LATTE '21, April 15, 2021, Virtual, Earth</mark>

## ACKNOWLEDGMENTS

We thank the reviewers for their feedback. This material is based upon work supported by the Defense Advanced Research Projects Agency (DARPA) under Contract No. HR0011-19-C-0106. Any opinions, findings, and conclusions or recommendations expressed in this material are those of the authors and do not necessarily reflect the views of DARPA.